\documentclass[
  reprint,
  prd,
  superscriptaddress,
  nofootinbib
]{revtex4-2}

\usepackage{graphicx}
\usepackage{amsmath}

\begin{document}

\title{Probing the spin--parity structure of hidden-charm pentaquarks
from spectroscopy and magnetic moments}

\author{Pallavi Gupta}
\email{pallavi.gupta@thapar.edu} 
\affiliation{Department of Physics and Materials Science,
Thapar Institute of Engineering and Technology, Patiala, India}

\date{\today}

\begin{abstract}
We investigate the spin-parity ($J^{P}$) assignments of experimentally observed hidden-charm pentaquark states within a baryon-meson molecular framework. The pentaquark mass spectrum is obtained using the G\"ursey--Radicati mass formula, with parameters fixed through a global fit to 41 experimentally established hadron masses. The resulting spectrum is then used to assign $J^{P}$ quantum numbers to the observed pentaquark candidates. Within this framework, the nonstrange states $P_{c}(4312)$, $P_{c}(4440)$, and $P_{c}(4457)$ are identified with the $J^{P}=1/2^{-}$, $3/2^{-}$, and $5/2^{-}$ configurations, respectively. The recently reported Belle state $P_{cs}(4459)$, which carries strangeness, is interpreted as the strange member of the SU(3) flavor octet with $J^{P}=3/2^{-}$. Magnetic moments are subsequently evaluated using explicitly constructed wave functions. Their systematic behavior across SU(3) flavor multiplets and different spin--parity assignments satisfies the expected sum-rule relations and indicates that magnetic moments can serve as a useful observable for refining the quantum-number identification of hidden-charm pentaquark states in future studies.

\end{abstract}

\maketitle

\section{Introduction}

The quark model proposed by Gell-Mann in 1964~\cite{GellMann1964} established SU(3) flavor symmetry as the foundation of hadron spectroscopy and naturally accommodates multiquark configurations beyond conventional mesons $(q\bar q)$ and baryons $(qqq)$. Among these possibilities, pentaquarks with quark content $qqqq\bar q$ have long been of theoretical interest. A major breakthrough occurred in 2015 when the LHCb Collaboration reported the first clear evidence for hidden-charm pentaquark candidates, $P_c(4380)$ and $P_c(4450)$~\cite{LHCb2015}, revitalizing the field and lending strong support to the hypothesis of exotic multiquark states.

Subsequent LHCb observations further expanded the spectrum of hidden-charm pentaquarks. In 2019, three additional structures—$P_c(4312)$, $P_c(4440)$, and $P_c(4457)$—were reported in the decay $\Lambda_b^0 \to J/\psi\, K^{-}$~\cite{Aaij2019}, confirming that the earlier $P_c(4450)$ signal consists of two narrower resonances. In 2020, the collaboration announced the first observation of a strange hidden-charm pentaquark, $P_{cs}(4459)$, reconstructed in $\Xi_b^- \to J/\psi\, \Lambda\, K^-$~\cite{Aaij2021}. This was followed by the observation of a new state, $P_c(4337)$, in 2021~\cite{Aaij2022}, and the strange charmed structure $P_{cs}(4338)$ in 2022~\cite{Aaij2023}. More recently, in 2025, the Belle Collaboration confirmed the existence of $P_{cs}(4459)$ with a slightly altered mass and width~\cite{Adachi2025}. The observed masses and decay widths of these states are summarized in Table\ref{tab:pentaquark_masses}.
\begin{table}[t]
\caption{Masses and decay widths of the observed pentaquark states.
Except for the last row (Belle collaboration), all data are from the LHCb collaboration.
All masses and decay widths are given in MeV.}
\begin{ruledtabular}
\begin{tabular}{ccc}
State & Mass & Decay width \\
\hline
$P_{c}(4380)$~\cite{LHCb2015} & $4380 \pm 8$ & $205 \pm 18$ \\
$P_{c}(4450)$~\cite{LHCb2015} & $4449.8 \pm 1.7$ & $39 \pm 5$ \\
$P_{c}(4312)$~\cite{Aaij2019} & $4311.9 \pm 0.7$ & $9.8 \pm 2.7$ \\
$P_{c}(4450)$~\cite{Aaij2019} & $4440.3 \pm 1.3$ & $20.6 \pm 4.9$ \\
$P_{c}(4457)$~\cite{Aaij2019} & $4457.3 \pm 0.6$ & $6.4 \pm 2.0$ \\
$P_{cs}(4459)$~\cite{Aaij2021} & $4458.8 \pm 2.9$ & $17.3 \pm 6.5$ \\
$P_{c}(4337)$~\cite{Aaij2022} & $4337 \pm 7$ & $29$ \\
$P_{cs}(4338)$~\cite{Aaij2023} & $4338.2 \pm 0.7$ & $7.0 \pm 1.2$ \\
$P_{cs}(4459)$~\cite{Adachi2025} & $4471.7 \pm 4.8$ & $22 \pm 13$ \\
\end{tabular}
\end{ruledtabular}
\label{tab:pentaquark_masses}
\end{table}

The rapid experimental progress in the past decade has stimulated significant theoretical efforts to determine the quantum numbers, internal structures, and underlying dynamics of the hidden-charm pentaquarks. 
The aforementioned pentaquark states have been investigated using a variety of theoretical frameworks, including effective field theories~\cite{Wang2024, Nakamura2023}, 
QCD sum rules~\cite{XWWang2023,Azizi2023}, machine learning–assisted analyses~\cite{Ferretti2022}, 
and quark model approaches~\cite{Yang2024,Ortega2023}. 
Beyond these methodologies, several structural interpretations have been proposed to explore the internal configurations of these exotic states, such as the hadronic molecular picture~\cite{He2019, Monemzadeh2016, Yang2017}, 
the diquark-diquark-antiquark model~\cite{Wang2021,Maiani2015,Wang2016}, 
the diquark-triquark model~\cite{Lebed2015,Zhu2016,Giron2021}, 
and the triangular singularity mechanism~\cite{Burns2023}.
However, the definitive determination of their spin-parity $J^P$ assignments remains incomplete. Among these proposals, the most favored picture is the molecular interpretation, in which a heavy baryon $(qqc)$ binds with a heavy-light meson $(q\bar c)$.
For hidden-charm pentaquarks containing two light quarks, SU(3) flavor symmetry allows the two light quarks in the heavy baryon to combine,
\[
3 \otimes 3 = 6 \oplus \bar{3},
\]
which then couples with the light quark from the meson to form
\[
6 \otimes 3 = 10 \oplus 8_{1f}, \qquad 
\bar{3} \otimes 3 = 1 \oplus 8_{2f}.
\]
These correspond to the decuplet, octet, and singlet flavor multiplets. 
The experimentally observed $P_c$ and $P_{cs}$ states are generally compatible
with an assignment to flavor-octet representations, although their spin-parity $J^P$ quantum numbers have not yet been firmly established. In this work, we extend our study of
hidden-charm molecular pentaquarks from the octet
to the decuplet by employing
the G\"ursey-Radicati (GR) \cite{GRmodel1,GRmodel2} mass framework to constrain the allowed $J^P$
assignments of the observed pentaquark candidates and to predict the masses of
their as yet unobserved partners in both the octet and decuplet multiplets. Since spectroscopy alone cannot fully determine the pentaquark structure, we further construct quark-model wavefunctions and use them to compute magnetic moments, hyperfine interactions, and mass splittings.
Our analysis of
magnetic moments reveal characteristic patterns associated with different
spin-parity assignments, charge states, and strangeness content, thereby
offering complementary structural information beyond mass spectra alone. The
combined study of spectroscopy and spin-dependent observables, thus provides a more robust framework for interpreting the nature of the observed $P_c$ and
$P_{cs}$ states, and for guiding future experimental searches for their predicted
partners.

 The paper is organized as follows. Section~2 presents the spectroscopy of hidden-charm pentaquarks and details the determination of GR model parameters using 41 experimentally established baryon masses \cite{PDG}, improving upon earlier analyses based on 21 inputs as done in Ref.\cite{PRL}—section~3 constructs wavefunctions for all relevant flavor multiplets $q\bar c qqc$. Section 4 employs these wavefunctions to calculate magnetic moments, followed by a summary in Section 5. 
\section{Mass Spectroscopy}
\label{sec:masses}

We used the Gürsey-Radicati (GR) mass operator,
which parametrizes baryon and multiquark masses in terms of spin, flavour, and
heavy-quark content.  The mass of a hadron belonging to an SU(3)$_F$ 
representation is written as
\begin{equation}
\begin{split}
M ={}&
M_0 + A\,S(S+1) + D\,Y \\
&+ E\!\left[I(I+1)-\frac{Y^2}{4}\right]
+ G\,C_2(\mathrm{SU}(3)) \\
&+ F_c\,N_c + F_b\,N_b .
\end{split}
\label{eq:GR}
\end{equation}

where $S$ is spin, $Y$ hypercharge, $I$ isospin, $C_2$ is the quadratic SU(3)
Casimir eigenvalue, and $N_c$, $N_b$ count the number of charm and bottom quarks.  
The coefficients $(M_0, A, D, E, G, F_c, F_b)$ are determined from a global fit
to experimentally available baryon masses.

To achieve a precise and stable determination of these coefficients, we perform a $\chi^2$ minimization fit to a comprehensive dataset of 41 experimentally established baryons, including the light octet and decuplet as well as singly charmed and singly bottom baryons. A uniform theoretical uncertainty of 0.5\% is assigned to each input mass to account for residual model dependence and to stabilize the fit. The complete list of baryon states used in the fit, together with the quantum numbers entering Eq.~(1), is summarized in Table~\ref{tab:baryon_inputs}.

The quality of the global fit can be quantified through the $\chi^2$ per degree of freedom. Using 41 baryon masses and seven free parameters $(M_0, A, D, E, G, F_c, F_b)$, the fit yields a total $\chi^2 = 82.16$ for 34 degrees of freedom, corresponding to $\chi^2/\text{d.o.f.} = 2.42$. Given the simplicity of the effective G\"ursey--Radicati mass operator and the wide mass range covered by the input states, this value indicates a satisfactory global description of the baryon spectrum.

The resulting best-fit values of the G\"ursey--Radicati mass parameters are
\begin{align*}
M_0 &= 964.76 \pm 6.76~\text{MeV}, \\
A   &= 20.59 \pm 2.34~\text{MeV}, \\
D   &= -191.54 \pm 2.15~\text{MeV}, \\
E   &= 33.78 \pm 1.63~\text{MeV}, \\
G   &= 45.34 \pm 2.58~\text{MeV}, \\
F_c &= 1366.76 \pm 4.80~\text{MeV}, \\
F_b &= 4836.01 \pm 10.94~\text{MeV},
\end{align*}
where the quoted uncertainties correspond to one standard deviation obtained from the covariance matrix of the fit.

These fitted parameters are then used to predict the masses of hidden-charm and hidden-bottom pentaquarks belonging to the SU(3)$_F$ octet and decuplet multiplets. The resulting mass spectra for different spin assignments are presented in Table~\ref{tab:penta_masses_ordered}, with the experimentally observed $P_c$ and $P_{cs}$ states highlighted for comparison.

\begin{table*}[t]
\caption{Baryon input states used in the GR mass-operator fit.
Masses are taken from the Particle Data Group and are given in MeV.}
\label{tab:baryon_inputs}
\begin{ruledtabular}
\begin{tabular}{lccccccc}
Baryon & Mass & $C_2$ & $S$ & $Y$ & $I$ & $N_c$ & $N_b$ \\
\hline
\multicolumn{8}{l}{\textbf{Light octet} ($J^P = \tfrac12^+$)} \\
$p$           & 938.272  & 3   & 1/2 & 1   & 1/2 & 0 & 0 \\
$n$           & 939.565  & 3   & 1/2 & 1   & 1/2 & 0 & 0 \\
$\Lambda$     & 1115.683 & 3   & 1/2 & 0   & 0   & 0 & 0 \\
$\Sigma^+$    & 1189.37  & 3   & 1/2 & 0   & 1   & 0 & 0 \\
$\Sigma^0$    & 1192.642 & 3   & 1/2 & 0   & 1   & 0 & 0 \\
$\Sigma^-$    & 1197.449 & 3   & 1/2 & 0   & 1   & 0 & 0 \\
$\Xi^0$       & 1314.86  & 3   & 1/2 & $-1$ & 1/2 & 0 & 0 \\
$\Xi^-$       & 1321.71  & 3   & 1/2 & $-1$ & 1/2 & 0 & 0 \\
\hline
\multicolumn{8}{l}{\textbf{Light decuplet} ($J^P = \tfrac32^+$)} \\
$\Delta^{++}$ & 1232.0 & 6 & 3/2 & 1   & 3/2 & 0 & 0 \\
$\Delta^{+}$  & 1232.0 & 6 & 3/2 & 1   & 3/2 & 0 & 0 \\
$\Delta^{0}$  & 1232.0 & 6 & 3/2 & 1   & 3/2 & 0 & 0 \\
$\Delta^{-}$  & 1232.0 & 6 & 3/2 & 1   & 3/2 & 0 & 0 \\
$\Sigma^{*+}$ & 1382.8 & 6 & 3/2 & 0   & 1   & 0 & 0 \\
$\Sigma^{*0}$ & 1383.7 & 6 & 3/2 & 0   & 1   & 0 & 0 \\
$\Sigma^{*-}$ & 1387.2 & 6 & 3/2 & 0   & 1   & 0 & 0 \\
$\Xi^{*0}$    & 1531.80& 6 & 3/2 & $-1$ & 1/2 & 0 & 0 \\
$\Xi^{*-}$    & 1535.0 & 6 & 3/2 & $-1$ & 1/2 & 0 & 0 \\
$\Omega^{-}$  & 1672.45& 6 & 3/2 & $-2$ & 0   & 0 & 0 \\
\hline
\multicolumn{8}{l}{\textbf{Charmed baryons} ($J^P = \tfrac12^+$)} \\
$\Lambda_c^+$      & 2286.46 & 4/3 & 1/2 &  2/3 & 0   & 1 & 0 \\
$\Sigma_c^{++}$    & 2453.97 & 10/3& 1/2 &  2/3 & 1   & 1 & 0 \\
$\Sigma_c^{+}$     & 2452.90 & 10/3& 1/2 &  2/3 & 1   & 1 & 0 \\
$\Sigma_c^{0}$     & 2453.75 & 10/3& 1/2 &  2/3 & 1   & 1 & 0 \\
$\Xi_c^{+}$        & 2467.93 & 4/3 & 1/2 & $-1/3$ & 1/2 & 1 & 0 \\
$\Xi_c^{0}$        & 2470.87 & 4/3 & 1/2 & $-1/3$ & 1/2 & 1 & 0 \\
$\Xi'_c{}^{+}$     & 2577.9  & 10/3& 1/2 & $-1/3$ & 1/2 & 1 & 0 \\
$\Xi'_c{}^{0}$     & 2578.8  & 10/3& 1/2 & $-1/3$ & 1/2 & 1 & 0 \\
$\Omega_c^0$       & 2695.2  & 10/3& 1/2 & $-4/3$ & 0   & 1 & 0 \\
\hline
\multicolumn{8}{l}{\textbf{Charmed baryons} ($J^P = \tfrac32^+$)} \\
$\Sigma_c^{*++}$   & 2518.41 & 10/3 & 3/2 & 2/3  & 1 & 1 & 0 \\
$\Sigma_c^{*+}$    & 2517.5  & 10/3 & 3/2 & 2/3  & 1 & 1 & 0 \\
$\Sigma_c^{*0}$    & 2518.48 & 10/3 & 3/2 & 2/3  & 1 & 1 & 0 \\
$\Xi_c^{*+}$       & 2645.53 & 10/3 & 3/2 & $-1/3$ & 1/2 & 1 & 0 \\
$\Xi_c^{*0}$       & 2646.32 & 10/3 & 3/2 & $-1/3$ & 1/2 & 1 & 0 \\
$\Omega_c^{*0}$    & 2765.9  & 10/3 & 3/2 & $-4/3$ & 0   & 1 & 0 \\
\hline
\multicolumn{8}{l}{\textbf{Bottom baryons} ($J^P = \tfrac12^+$)} \\
$\Lambda_b^0$      & 5619.60 & 4/3 & 1/2 & 4/3 & 0   & 0 & 1 \\
$\Xi_b^0$          & 5791.9  & 4/3 & 1/2 & 1/3 & 1/2 & 0 & 1 \\
$\Xi_b^{-}$        & 5797.0  & 4/3 & 1/2 & 1/3 & 1/2 & 0 & 1 \\
$\Sigma_b^{+}$     & 5811.3  & 10/3& 1/2 & 4/3 & 1   & 0 & 1 \\
$\Sigma_b^{-}$     & 5815.5  & 10/3& 1/2 & 4/3 & 1   & 0 & 1 \\
$\Omega_b^{-}$     & 6046.1  & 10/3& 1/2 & $-2/3$ & 0 & 0 & 1 \\
\hline
\multicolumn{8}{l}{\textbf{Bottom baryons} ($J^P = \tfrac32^+$)} \\
$\Sigma_b^{*+}$    & 5832.1 & 10/3 & 3/2 & 4/3 & 1 & 0 & 1 \\
$\Sigma_b^{*-}$    & 5836.5 & 10/3 & 3/2 & 4/3 & 1 & 0 & 1 \\
\end{tabular}
\end{ruledtabular}
\end{table*}

\begin{table*}[t]
\caption{Predicted masses of hidden-charm and hidden-bottom pentaquarks
(in MeV) in the SU(3)$_F$ octet and decuplet multiplets, ordered by spin $J$.
Experimentally observed states matched to the predictions are shown in bold.}
\label{tab:penta_masses_ordered}
\begin{ruledtabular}
\begin{tabular}{cccccccc}
State & Multiplet & $Y$ & $I$ & $J$ & $M_{c\bar c}$ & $M_{b\bar b}$ & Experiment \\
\hline
\multicolumn{8}{c}{\textbf{Octet} $(8)$} \\
\hline
\multicolumn{8}{l}{Spin $J=\tfrac12$} \\
$P_{Q}$     & 8 & 1    & 1/2 & 1/2 & \textbf{4318.29} & 11256.78 & \textbf{$P_c(4312)$} \\
$P_{QS}$    & 8 & 0    & 0   & 1/2 & \textbf{4492.93} & 11391.41 & \textbf{$P_{cs}(4338)$} \\
$P_{QSS}$   & 8 & $-1$ & 1/2 & 1/2 & 4701.36 & 11599.84 & -- \\
\hline
\multicolumn{8}{l}{Spin $J=\tfrac32$} \\
$P_{Q}$     & 8 & 1    & 1/2 & 3/2 & \textbf{4380.05} & 11318.55 & \textbf{$P_c(4380)$} \\
$P_{QS}$    & 8 & 0    & 0   & 3/2 & \textbf{4554.70} & 11453.18 & \textbf{$P_{cs}(4458)$} \\
$P_{QSS}$   & 8 & $-1$ & 1/2 & 3/2 & 4763.12 & 11661.60 & -- \\
\hline
\multicolumn{8}{l}{Spin $J=\tfrac52$} \\
$P_{Q}$     & 8 & 1    & 1/2 & 5/2 & \textbf{4483.00} & 11421.49 & \textbf{$P_c(4440/4457)$} \\
$P_{QS}$    & 8 & 0    & 0   & 5/2 & 4657.64 & 11556.12 & -- \\
$P_{QSS}$   & 8 & $-1$ & 1/2 & 5/2 & 4866.07 & 11764.55 & -- \\
\hline
\multicolumn{8}{c}{\textbf{Decuplet} $(10)$} \\
\hline
\multicolumn{8}{l}{Spin $J=\tfrac12$} \\
$P_{Q}$     & 10 & 1    & 3/2 & 1/2 & 4555.67 & 11494.17 & -- \\
$P_{QS}$    & 10 & 0    & 1   & 1/2 & 4696.53 & 11635.03 & -- \\
$P_{QSS}$   & 10 & $-1$ & 1/2 & 1/2 & 4837.39 & 11775.88 & -- \\
$P_{QSSS}$  & 10 & $-2$ & 0   & 1/2 & 4978.25 & 11916.74 & -- \\
\hline
\multicolumn{8}{l}{Spin $J=\tfrac32$} \\
$P_{Q}$     & 10 & 1    & 3/2 & 3/2 & 4617.44 & 11555.93 & -- \\
$P_{QS}$    & 10 & 0    & 1   & 3/2 & 4758.30 & 11696.79 & -- \\
$P_{QSS}$   & 10 & $-1$ & 1/2 & 3/2 & 4899.16 & 11837.65 & -- \\
$P_{QSSS}$  & 10 & $-2$ & 0   & 3/2 & 5040.02 & 11978.51 & -- \\
\hline
\multicolumn{8}{l}{Spin $J=\tfrac52$} \\
$P_{Q}$     & 10 & 1    & 3/2 & 5/2 & 4720.38 & 11658.88 & -- \\
$P_{QS}$    & 10 & 0    & 1   & 5/2 & 4861.24 & 11799.74 & -- \\
$P_{QSS}$   & 10 & $-1$ & 1/2 & 5/2 & 5002.10 & 11940.60 & -- \\
$P_{QSSS}$  & 10 & $-2$ & 0   & 5/2 & 5142.96 & 12081.46 & -- \\
\end{tabular}
\end{ruledtabular}
\end{table*}

\begin{figure}[t]
\centering
\includegraphics[width=0.9\columnwidth]{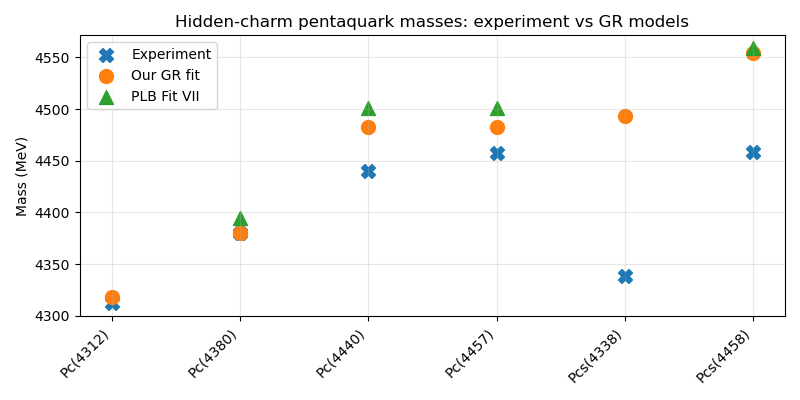}
\caption{Comparison of GR-predicted pentaquark masses with experimental LHCb values
and earlier results from Ref.~\cite{PRL}.}
\label{fig:mu_fig1}
\end{figure}

SU(3)$_F$ mass hierarchy of the predicted pentaquark charm octet, shows excellent agreement 
with the N-$\Lambda$-$\Xi$ pattern of the baryon octet. Figures~\ref{fig:mu_fig1} show, 
respectively, the comparison of our predicted masses with the experiment and masses predicted in Ref.\cite{PRL}. The closeness of our calculated masses to the experimental data indicates the improved accuracy of our fitted parameters. 

Since most pentaquarks lack experimentally determined $J^P$ values, we assign
spin by matching each experimental mass to the closest predicted octet state
in Table~\ref{tab:penta_masses_ordered}.  
This yields the following identifications:
\begin{itemize}
    \item $\mathrm{P_c}(4312)$ as $J=1/2$ ($Y=1, I=1/2$),
    \item $\mathrm{P_c}(4380)$ as $J=3/2$ ($Y=1$, I=1/2),
    \item $\mathrm{P_c}(4440)$ and $\mathrm{P_c}(4457)$ as $J=5/2$ ($Y=1, I=1/2$),
    \item $\mathrm{P_{cs}}(4338)$ as $J=1/2$ ($Y=0, I=1$),
    \item $\mathrm{P
    _{cs}}(4458)$ as $J=3/2$ ($Y=0, I=1$).
\end{itemize}
The stability of these identifications under variation of the GR coefficients 
demonstrates the reliability of the assigned $J^P$ values. These assignments are driven purely by mass proximity and therefore represent the 
most conservative interpretation of the LHCb data within the GR framework.  
Their stability reflects the internal consistency of the fitted operator and supports the identification of the $J^P$ values proposed above.

The mass spectroscopy presented in this section shows that fitting the GR mass
operator to a large, diverse dataset of 41 baryons yields a parameter set with
remarkable predictive power.  
The predicted hidden-charm and hidden-bottom pentaquark masses
(Table~\ref{tab:penta_masses_ordered}) agree closely with experimental $P_c$ and 
$P_{cs}$ states and improve upon earlier Ref.\cite{PRL} results based on a smaller baryon dataset. Notably, for the two highest-mass states shown in Fig.\ref{fig:mu_fig1}, including the strange $P_{cs}$ sector, the difference between our predicted masses and the experimental values is at the level of only a few percent, typically in the range of $2\text{-}4\%$. Deviations of this size are expected for an effective mass operator constrained exclusively by baryon spectroscopy and indicate that the present fit successfully reproduces the overall mass scale and hierarchy of the pentaquark spectrum. These residual discrepancies are likely to be further reduced in future studies as additional pentaquark states become experimentally established and can be incorporated into an extended fitting framework.

To further validate the values of the fitted mass parameters, we 
performed the following three independent mass-splitting analyses.  Splittings provide stringent 
and largely model-independent diagnostics: if the GR operator captures the correct 
underlying QCD behavior, then the splittings in the pentaquark sector should mirror
those seen in ordinary baryons. 

\begin{itemize}
  \item Heavy Quark Splitting\\Using masses listed in PDF\cite{PDG}, heavy quark splittings in baryons are
  \begin{align*}
  M_b^{\Lambda}-M_c^{\Lambda} =  3333.1 ~{\rm MeV},\\
  M_b^{\Sigma}-M_c^{\Sigma} =  3359.9 ~{\rm MeV}, \\
  M_b^{\Xi}-M_c^{\Xi} =  3325.0 ~{\rm MeV}.
  \end{align*}
  Thus, ordinary baryons exhibit a nearly universal heavy-quark mass displacement of
$\Delta M^{\rm baryon}_{b-c}  \approx (3.33 \pm 0.03)\ \mathrm{GeV}$,
independent of their SU(3)$_F$ quantum numbers.  
Since hidden-charm and hidden-bottom pentaquarks contain a heavy quark and a 
heavy antiquark, so this splitting should be approximately twice this shift
$\Delta M^{\rm penta}_{b-c} \approx 2\,\Delta M^{\rm baryon}_{b-c}
\approx 6.7~\mathrm{GeV}$.
Using the charm and bottom pentaquark masses calculated by our GR parameters, we get $\Delta M^{\rm penta}_{b-c} = 6.91 \pm 0.02 \ ~{\rm GeV}$ for both octet and decuplet states. The results clearly demonstrate that 
$\Delta M^{\rm penta}_{b-c} \approx 2\,\Delta M^{\rm baryon}_{b-c}$, 
as expected for states containing a $Q\bar Q$ pair thus authenticating the fitted parameters and thus assigned $J^{P}$'s to experimental states. 

  \item SU(3)$_F$ flavour Splitting
  \\For the baryon octet, SU(3)$_F$ flavour symmetry breaking produces the characteristic hierarchy
\begin{equation*}
\begin{aligned}
\Delta_{1,0}^{\rm baryon} &= M(Y=0) - M(Y=1) \approx 176.8~{\rm MeV},\\
\Delta_{0,-1}^{\rm baryon} &= M(Y=-1) - M(Y=0) \approx 202.6~{\rm MeV}.
\end{aligned}
\end{equation*}
Remarkably, the pentaquark octet predicted by our GR operator reproduces these 
splittings with differences of only $2$-$5$~MeV:

\begin{align*}
\Delta_{1,0}^{\rm penta} &= 174.6~{\rm MeV},\qquad
\Delta_{0,-1}^{\rm penta} &= 208.4~{\rm MeV}.
\end{align*}

The numerical proximity to baryon splittings provides strong evidence that the 
SU(3)$_F$ breaking terms extracted from the 41-baryon fit remain valid when applied 
to multiquark systems.  
\item Spin Splittings \\Spin-dependent mass differences offer a final, independent validation.  
For charmed baryons, heavy-quark spin symmetry yields the well-known splitting.
\begin{align*}
M(\Sigma_c^*(S=3/2)) - M(\Sigma_c(S=1/2)) = 64.59 \pm 0.14 ~{\rm MeV} \\
M(\Xi_c^*(S=3/2)) - M(\Xi_c(S=1/2)) = 67.57 \pm 0.06 ~{\rm MeV} 
\end{align*}

Our predicted charm and bottom pentaquark octet masses give
\begin{equation*}
\begin{aligned}
\Delta^{(c\bar c)}(3/2 - 1/2) \approx 61.74 \pm 0.03 ~{\rm MeV},\\
\Delta^{(c\bar c)}(5/2 - 3/2) \approx 102.94 \pm 0.01 ~{\rm MeV}\\
\Delta^{(b\bar b)}(3/2 - 1/2) \approx 61.77 \pm 0.01 ~{\rm MeV}, \\
\Delta^{(b\bar b)}(5/2 - 3/2) \approx 102.94 \pm 0.01 ~{\rm MeV}
\end{aligned}
\end{equation*}
The results are in intriguing agreement with the charmed baryon value.  
Because the GR operator contains no explicit $1/m_Q$ spin-suppression term, the 
bottom pentaquark splittings are identical to those of the charm sector; these  
constitute definite model predictions awaiting experimental confirmation. 
\end{itemize}

Heavy-quark $(b-c)$ splittings, SU(3)$_F$ hypercharge splittings, and 
spin-dependent splittings derived reproduce the 
corresponding baryon patterns with high precision. The simultaneous agreement of absolute masses and internal splitting structures 
provides strong evidence that the fitted GR operator effectively encodes the 
essential QCD dynamics governing multiquark systems. Consequently, the predicted 
pentaquark masses carry small theoretical uncertainties, lending confidence to 
the corresponding $J^P$ assignments and to the hidden-bottom predictions proposed 
here as robust benchmarks for future experimental investigations.

However, mass spectroscopy alone cannot fully reveal the internal structure of 
pentaquarks. To probe their underlying dynamics more deeply, it is necessary to 
analyze their electromagnetic properties, which requires explicit construction of 
the pentaquark wave functions. In the next section, we therefore develop the wave 
functions of the hidden-charm pentaquark octet and decuplet multiplets.

\section{Wavefunctions}
In the literature, several internal configurations have been proposed for 
describing the five-quark structure of pentaquarks. 
The molecular picture treats the pentaquark $(qqqq\bar{q})$ as a loosely 
bound hadronic system composed of a baryon $(qqq)$ and a meson $(q\bar{q})$. 
In contrast, the diquark-diquark-antiquark model assumes a more compact 
structure, where the pentaquark is formed from two correlated diquarks $(qq)$, $(qq)$ 
and an antiquark $(\bar{q})$. 
A third possibility, the diquark-triquark model, interprets the state as arising 
from the clustering of a diquark $(qq)$ and a triquark $(qq\bar{q})$.

Among these descriptions, the molecular model is the most extensively used and has 
successfully reproduced several features of the observed hidden-charm pentaquark 
candidates. Motivated by this, in the present section, we adopt the molecular picture as 
our working framework and construct the explicit flavor-spin wave functions of the 
hidden-charm pentaquark states belonging to the SU(3) flavor octet and decuplet 
multiplets. These wave functions subsequently serve as essential inputs for evaluating 
electromagnetic properties, particularly the magnetic moments, which provide deeper 
insight into the internal structure of the pentaquark states.

The wavefunction $\Psi$ of a hadron is composed of four degrees of freedom-spin, color, flavor, and space. The general representation is 
\begin{equation}
    \Psi = \psi_{flavour}\chi_{spin}\eta_{space}\xi_{color}
\end{equation}

 The pentaquarks are fermionic multiquark systems, and the total wave function must be antisymmetric under the exchange of identical quarks. In the molecular picture, both the baryon and meson constituents are individually color singlets; consequently, their bound state is automatically a color-singlet configuration, and no additional color antisymmetrization between quarks belonging to different clusters is required. For ground-state pentaquarks with vanishing relative orbital angular momentum, the spatial part of the wave function is symmetric, and therefore the required antisymmetry is ensured by an appropriate symmetry of the spin–flavor wave function. 
 In the hidden charm pentaquark molecular state (q$\bar{c}$)(qqc), the two light quarks of the baryons can be symmetric or antisymmetric $
3 \otimes 3 = 6_{\text{sym}} + \bar{3}_{\text{antisym}}$.
In the former case, $6_{fs}$ couples with $3_f$ flavor representation of light quark of heavy charm meson to form $6_f \otimes 3_f = 10_f + 8_1f$. In the latter case, antisymmetric flavor $\bar3_{fa}$ combines with $3_f$ of light quark of the meson to give $\bar3_{fa} \otimes 3_f = 1 + 8_2f$. Therefore, the hidden-charm pentaquark states form both decuplet and octet representations. Including the charm quark ($c$), its antiparticle ($\bar{c}$), and the corresponding Clebsch–Gordan coefficients, we construct the flavor wave functions for the decuplet and the two octets in the $(\bar{c}q_1)(cq_2q_3)$ molecular configuration. These wave functions are listed in Table~\ref{tab:hc_octet_final} and Table~\ref{tab:hc_decuplet_final}.
\begin{table*}[t]
\caption{Flavor wave functions of hidden-charm pentaquark octet states.
For each state, the first row corresponds to the symmetric flavor octet
$8_{1f}$ with diquark $\{q_2 q_3\}$, and the second row to the antisymmetric
octet $8_{2f}$ with diquark $[q_2 q_3]$. The quantum numbers $Y$, $I$, and
$I_3$ denotes hypercharge, isospin, and its third component, respectively.}
\label{tab:hc_octet_final}
\begin{ruledtabular}
\begin{tabular}{cccc}
State & Flavor & $(Y,I,I_3)$ & Wave function \\
\hline
$P_{c}^{+}$ 
& $8_{1f}$ & $(1,\tfrac12,\tfrac12)$ &
$\sqrt{\tfrac{2}{3}}\,d\bar c\,\{uu\}c
- \sqrt{\tfrac{1}{3}}\,u\bar c\,\{ud\}c$ \\
& $8_{2f}$ &  &
$u\bar c\,[ud]c$ \\
\hline
$P_{c}^{0}$ 
& $8_{1f}$ & $(1,\tfrac12,-\tfrac12)$ &
$\tfrac{1}{\sqrt3}\,d\bar c\,\{ud\}c
- \sqrt{\tfrac{2}{3}}\,u\bar c\,\{dd\}c$ \\
& $8_{2f}$ &  &
$d\bar c\,[ud]c$ \\
\hline
$P_{cs}^{+}$ 
& $8_{1f}$ & $(0,1,1)$ &
$\tfrac{1}{\sqrt3}\,u\bar c\,\{us\}c
- \sqrt{\tfrac{2}{3}}\,s\bar c\,\{uu\}c$ \\
& $8_{2f}$ &  &
$u\bar c\,[us]c$ \\
\hline
$P_{cs}^{0}$ 
& $8_{1f}$ & $(0,1,0)$ &
$\tfrac{1}{\sqrt6}\,d\bar c\,\{us\}c
+ \tfrac{1}{\sqrt6}\,u\bar c\,\{ds\}c
- \sqrt{\tfrac{2}{3}}\,s\bar c\,\{ud\}c$ \\
& $8_{2f}$ &  &
$\tfrac{1}{\sqrt2}
\big(d\bar c\,[us]c + u\bar c\,[ds]c\big)$ \\
\hline
$P_{cs}^{-}$ 
& $8_{1f}$ & $(0,1,-1)$ &
$\tfrac{1}{\sqrt3}\,d\bar c\,\{ds\}c
- \sqrt{\tfrac{2}{3}}\,s\bar c\,\{dd\}c$ \\
& $8_{2f}$ &  &
$d\bar c\,[ds]c$ \\
\hline
$P_{cs}^{0}$ 
& $8_{1f}$ & $(0,0,0)$ &
$\tfrac{1}{\sqrt2}
\big(u\bar c\,\{ds\}c - d\bar c\,\{us\}c\big)$ \\
& $8_{2f}$ &  &
$\tfrac{1}{\sqrt6}
\big(d\bar c\,[us]c - u\bar c\,[ds]c
- 2s\bar c\,[ud]c\big)$ \\
\hline
$P_{css}^{+}$ 
& $8_{1f}$ & $(-1,\tfrac12,\tfrac12)$ &
$\tfrac{1}{\sqrt3}\,s\bar c\,\{us\}c
- \sqrt{\tfrac{2}{3}}\,u\bar c\,\{ss\}c$ \\
& $8_{2f}$ &  &
$s\bar c\,[us]c$ \\
\hline
$P_{css}^{0}$ 
& $8_{1f}$ & $(-1,\tfrac12,-\tfrac12)$ &
$\tfrac{1}{\sqrt3}\,s\bar c\,\{ds\}c
- \sqrt{\tfrac{2}{3}}\,d\bar c\,\{ss\}c$ \\
& $8_{2f}$ &  &
$s\bar c\,[ds]c$ \\
\end{tabular}
\end{ruledtabular}
\end{table*}

\begin{table*}[t]
\caption{Flavor wave functions of hidden-charm pentaquark decuplet states.
The quantum numbers $Y$, $I$, and $I_3$ denote hypercharge, isospin, and its
third component, respectively.}
\label{tab:hc_decuplet_final}
\begin{ruledtabular}
\begin{tabular}{ccc}
State & $(Y,I,I_3)$ & Wave function \\
\hline
$P^{++}_{c}$ 
& $(1,\tfrac{3}{2},\tfrac{3}{2})$ &
$u\bar c\,\{uu\}c$ \\
\hline
$P^{+}_{c}$ 
& $(1,\tfrac{3}{2},\tfrac{1}{2})$ &
$\tfrac{1}{\sqrt3}\, d\bar c\,\{uu\}c
+ \sqrt{\tfrac{2}{3}}\, u\bar c\,\{ud\}c$ \\
\hline
$P^{0}_{c}$ 
& $(1,\tfrac{3}{2},-\tfrac{1}{2})$ &
$\tfrac{1}{\sqrt3}\, u\bar c\,\{dd\}c
+ \sqrt{\tfrac{2}{3}}\, d\bar c\,\{ud\}c$ \\
\hline
$P^{-}_{c}$ 
& $(1,\tfrac{3}{2},-\tfrac{3}{2})$ &
$d\bar c\,\{dd\}c$ \\
\hline
$P^{+}_{cs}$ 
& $(0,1,1)$ &
$\tfrac{1}{\sqrt3}\, s\bar c\,\{uu\}c
+ \sqrt{\tfrac{2}{3}}\, u\bar c\,\{us\}c$ \\
\hline
$P^{0}_{cs}$ 
& $(0,1,0)$ &
$\tfrac{1}{\sqrt3}
\big(s\bar c\,\{ud\}c + d\bar c\,\{us\}c + u\bar c\,\{ds\}c\big)$ \\
\hline
$P^{-}_{cs}$ 
& $(0,1,-1)$ &
$\tfrac{1}{\sqrt3}\, s\bar c\,\{dd\}c
+ \sqrt{\tfrac{2}{3}}\, d\bar c\,\{ds\}c$ \\
\hline
$P^{0}_{css}$ 
& $(-1,\tfrac{1}{2},\tfrac{1}{2})$ &
$\tfrac{1}{\sqrt3}\, u\bar c\,\{ss\}c
+ \sqrt{\tfrac{2}{3}}\, s\bar c\,\{us\}c$ \\
\hline
$P^{-}_{css}$ 
& $(-1,\tfrac{1}{2},-\tfrac{1}{2})$ &
$\tfrac{1}{\sqrt3}\, d\bar c\,\{ss\}c
+ \sqrt{\tfrac{2}{3}}\, s\bar c\,\{ds\}c$ \\
\hline
$P^{-}_{csss}$ 
& $(-2,0,0)$ &
$s\bar c\,\{ss\}c$ \\
\end{tabular}
\end{ruledtabular}
\end{table*}

\section{Magnetic Moments of Ground-State Hidden-Charm Pentaquark States}

In the constituent quark model, the magnetic moment of a multiquark system originates from the intrinsic spin of its constituent quarks. The corresponding magnetic moment operator is given by
\begin{equation*}
\hat{\mu} = \sum_i \frac{q_i}{2m_i}\,\boldsymbol{\sigma}_i ,
\label{eq:mu_operator}
\end{equation*}
where $q_i$, $m_i$, and $\sigma_i$ denote the electric charge, constituent mass, and Pauli spin operator of the $i$-th quark, respectively.

In the molecular picture adopted in this work, the hidden-charm pentaquark is described as a baryon-meson bound state. For the ground state, the relative orbital angular momentum between the baryon and meson clusters is taken to be $L=0$. Consequently, the orbital contribution to the magnetic moment vanishes identically, and the total magnetic moment operator reduces to
\begin{equation*}
\hat{\mu} = \hat{\mu}_B + \hat{\mu}_M ,
\label{eq:total_mu}
\end{equation*}
where $\hat{\mu}_B$ and $\hat{\mu}_M$ denote the intrinsic magnetic moments of the baryon and meson constituents, respectively.

The intrinsic magnetic moment of the baryon constituent is expressed as
\begin{equation*}
\hat{\mu}_B = \sum_{i=1}^{3} \mu_i\,\sigma_i ,
\label{eq:mu_baryon}
\end{equation*}
while the mesonic contribution takes the form
\begin{equation*}
\hat{\mu}_M = \sum_{i=1}^{2} \mu_i\,\sigma_i ,
\label{eq:mu_meson}
\end{equation*}
where $\mu_i = q_i/(2m_i)$ represents the magnetic moment of the corresponding constituent quark or antiquark.
The magnetic moment of a ground-state hidden-charm pentaquark $(\bar{c}q_1)(cq_2q_3)$  is obtained by evaluating the expectation value of the total magnetic moment operator,
\begin{equation*}
\mu = \langle \psi \,|\, \hat{\mu}_B + \hat{\mu}_M \,|\, \psi \rangle ,
\label{eq:mu_expectation}
\end{equation*}
where $\psi$ denotes the full pentaquark wave function.
After explicitly performing the spin couplings, the magnetic moment can be written in terms of Clebsch-Gordan coefficients \cite{Mutuk:2024jxf},\cite{mmop} as

\begin{widetext}
\begin{align}
\mu &=
\sum_{S S_z,\, l l_z}
\left\langle S S_z,\, l l_z \middle| J J_z \right\rangle^2
\Bigg\{
\sum_{S_B',\, S_M'}
\left\langle S_B S_B',\, S_M S_M' \middle| S S_z \right\rangle^2
\times
\Bigg[
S_M'(\mu_c + \mu_{q_1})
+ \sum_{S_c'}
\left\langle
S_c S_c',\, S_D (S_B' - S_c')
\middle| S_B S_B'
\right\rangle^2
\nonumber\\[1mm]
&\qquad\qquad\qquad\qquad\qquad\qquad\times
\Big(
g\,\mu_c\,S_c'
+ (S_B' - S_c')(\mu_{q_2} + \mu_{q_3})
\Big)
\Bigg]
\Bigg\}.
\label{eq:mu_final}
\end{align}
\end{widetext}

Here, $\psi$ denotes the pentaquark wave function listed in Tables~\ref{tab:hc_octet_final} and~\ref{tab:hc_decuplet_final}. The total spin $S$ of the pentaquark is obtained by coupling the spin of the baryon constituent $S_B$ with that of the meson constituent $S_M$. The corresponding Clebsch--Gordan coefficients account for the different spin configurations arising from $S_B \otimes S_M \to S$, which are then combined with the orbital contribution ($L = 0$) to form the total spin $J$. This decomposition allows us to systematically evaluate the magnetic-moment contributions from the baryon and meson sectors for each allowed total spin assignment.

The quantity $S_D$ represents the spin of the diquark subsystem inside the baryon, while the primed symbols denote intermediate spin projections arising from angular-momentum coupling. The parameters $\mu_c$, $\mu_{q_1}$, $\mu_{q_2}$, and $\mu_{q_3}$ denote the magnetic moments of the charm and light constituent quarks, and $g$ is the gyromagnetic factor associated with the heavy quark.

We use the constituent quark masses from Ref\cite{mass}:
$m_u = 0.338~\mathrm{GeV}$,
$m_d = 0.350~\mathrm{GeV}$,
$m_s = 0.500~\mathrm{GeV}$,
and $m_c = 1.275~\mathrm{GeV}$.
We calculate the magnetic moments of hidden-charm pentaquark states for three possible spin-parity assignments:
$J^P = \frac{1}{2}^-$, $\frac{3}{2}^-$, and $\frac{5}{2}^-$.

The total spin-parity of a pentaquark state is defined as
\begin{equation*}
J^P = J_B^{P_B} \otimes J_M^{P_M} \otimes J_L^{P_L},
\end{equation*}
where $J_B^{P_B}$ and $J_M^{P_M}$ denote the spin and parity of the baryon and meson constituents, respectively, and $J_L^{P_L}$ represents the total orbital angular momentum and parity associated with their relative motion. For the ground-state pentaquarks considered in this work, the relative orbital angular momentum between the baryon and meson is taken to be $L = 0$. The numerical values of the magnetic moments obtained for different $J^P$ configurations are summarized in Tables~\ref{tab:decuplet_magnetic_moments}, ~\ref{tab:octet1f_magnetic_moments}, ~\ref{tab:octet2f_magnetic_moments} for the decuplet, symmetric octet ($8_{1f}$), and antisymmetric octet ($8_{2f}$) representations, respectively.

\begin{figure}[t]
\centering
\includegraphics[width=0.9\columnwidth]{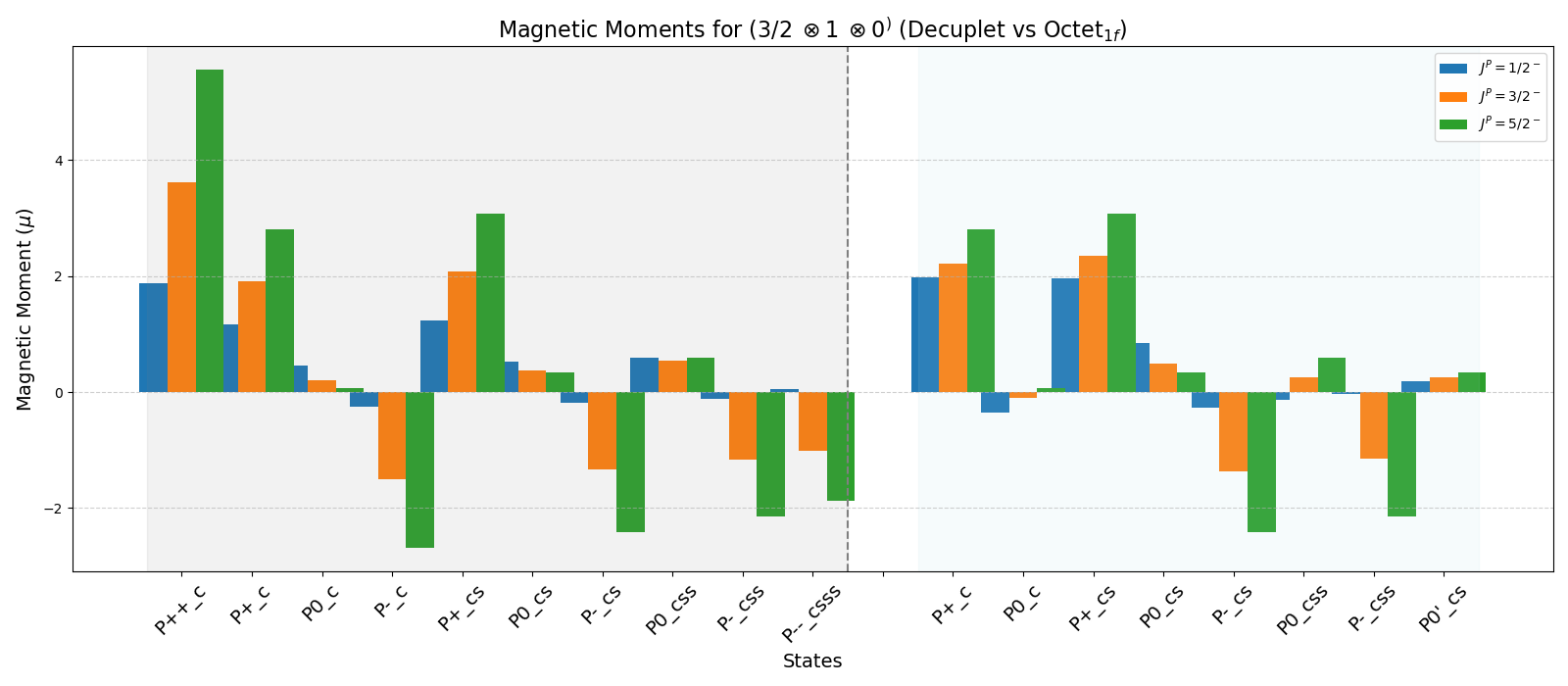}
\caption{Magnetic moments of hidden-charm pentaquark states for different
spin--parity assignments, comparing the decuplet $10_f$ and octet $8_{1f}$
flavor representations. }
\label{fig:2}
\end{figure}
\begin{figure}[t]
\centering
\includegraphics[width=0.9\columnwidth]{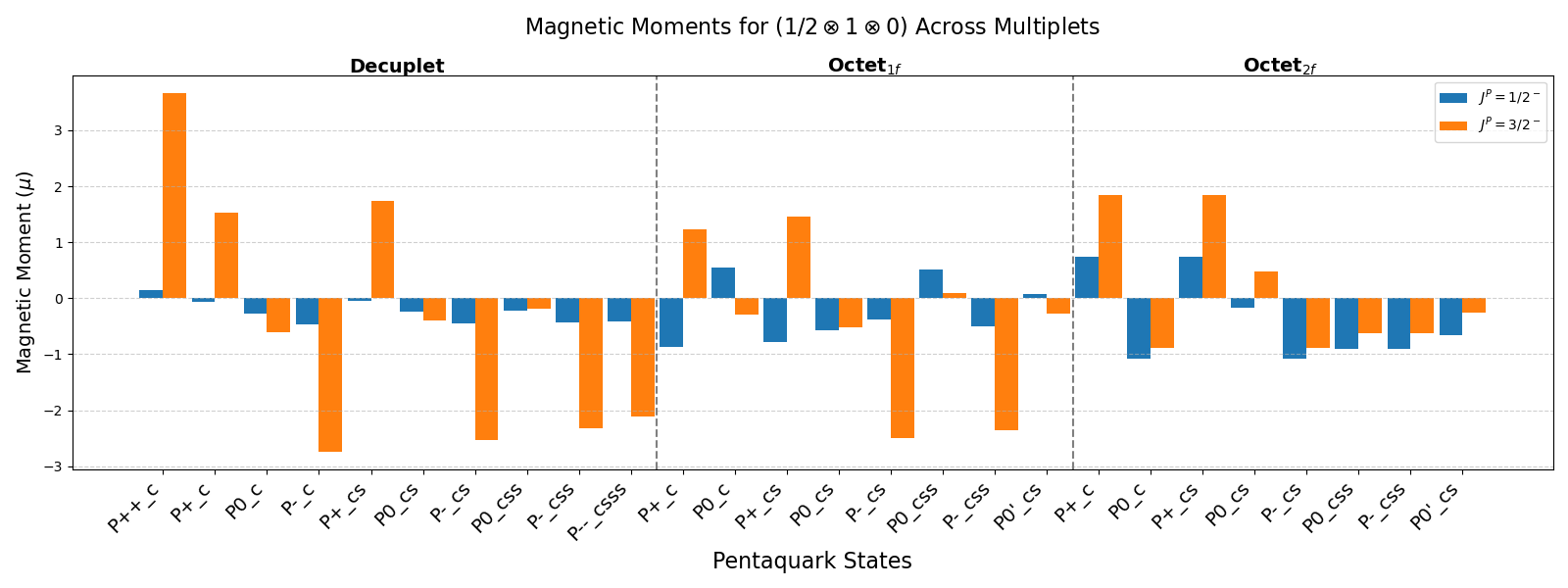}
\caption{Magnetic moments of hidden-charm pentaquark states as functions of
electric charge and strangeness for different spin--parity assignments.}
\label{fig:3}
\end{figure}

To elucidate the behavior of magnetic moments across different multiplets, we present the results in Figures \ref{fig:2} and \ref{fig:3}. 
Figure \ref{fig:2} provides a comparative analysis of the magnetic moments for states with $J^P = \tfrac{1}{2}^-$, $\tfrac{3}{2}^-$, and $\tfrac{5}{2}^-$, derived from the $\tfrac{3}{2}^+ \!\otimes \! 1^- \!\otimes \!0^+$ coupling in baryon–meson systems at the ground state. This analysis encompasses both the decuplet ($\mathbf{10}_f$) and the symmetric octet ($\mathbf{8}_1$) multiplets. 
Figure\ref{fig:3} shows the dependence of magnetic moments on electric charge and strangeness for $J^P = \tfrac{1}{2}^-$ and $\tfrac{3}{2}^-$ states, resulting from the  $\tfrac{1}{2}^+ \otimes 1^- \otimes 0^+$
 coupling across the $\mathbf{10}_f$, $\mathbf{8}_1$, and $\mathbf{8}_2$ multiplets. The principal findings are summarized as follows:

\begin{itemize}

\item \textbf{Charge dependence and neutral-state behavior:}  
The magnetic moments show a systematic dependence on the electric charge of the pentaquark states. Positively charged states generally carry large positive magnetic moments, while negatively charged states exhibit increasingly negative values, with the largest magnitudes occurring for the highest spin assignments. In contrast, electrically neutral pentaquark states display magnetic moments that remain close to zero and exhibit only a weak dependence on the spin, indicating substantial cancellations among the constituent-quark contributions.

\item \textbf{Spin enhancement:}  
For a fixed flavor composition, the magnitude of the magnetic moment increases with the total spin of the state, following the ordering
\[
|\mu(J = 5/2)| > |\mu(J = 3/2)| > |\mu(J = 1/2)| .
\]
This behavior reflects the enhanced alignment of the baryon--meson constituent spins at higher total spin and confirms the internal consistency of the angular-momentum coupling scheme used in the construction of the pentaquark wave functions.

\item \textbf{Strangeness suppression:}  
A systematic suppression of the magnetic-moment magnitude is observed with increasing strangeness content. As one moves from non-strange $P_c$ states to singly and doubly strange $P_{cs}$ and $P_{css}$ configurations, the magnetic moments decrease regularly. This trend originates from the larger constituent mass and reduced magnetic moment of the strange quark and persists across all considered spin assignments.

\item \textbf{Multiplet discrimination:}  
For identical charge and spin quantum numbers, decuplet states typically exhibit larger magnetic moments than their octet counterparts, while the two octet representations show distinct patterns. This separation highlights the sensitivity of magnetic moments to the underlying SU(3)$_F$ flavor structure and demonstrates their usefulness as a discriminating observable between different multiplet assignments.

  \item \textbf{Isospin Symmetry:} Under the assumption of isospin symmetry ($m_u=m_d$), the magnetic moment of states with isospin projection $I_3=-\tfrac12$ vanishes. This follows directly from the analytical expression $\mu = 2\mu_d + \mu_u$,
 which becomes zero when $\mu_u=-2\mu_d$. This cancellation explains the near-zero magnetic moments obtained for several neutral configurations in the numerical analysis.

  \item \textbf{Octet $8_{2f}$:} For the octet $8_{2f}$ pentaquarks with $J^P=\tfrac12^-$, the first configuration $(\tfrac12^+ \otimes 0^- \otimes 0^+)$ leads to a magnetic moment that depends solely on the charm-quark contribution, $\mu = \mu_c $. Consequently, all eight states acquire identical magnetic moments, independent of charge and strangeness. This exact degeneracy is clearly reflected in Table\ref{tab:octet2f_magnetic_moments} and corroborated by the flat pattern observed in Fig \ref{fig:3}. For the second configuration $(\tfrac12^+ \otimes 1^- \otimes 0^+)$ in the $8_{2f}$ multiplet, the general expression for the magnetic moment is $ \mu = -\mu_c + \frac{2}{3}\mu_{q_1}$, where $\mu_{q_1}$ denotes the magnetic moment of the light constituent quark in the meson. As a result, the magnetic moments are identical for states related by isospin symmetry, such as $(I,I_3)=(\tfrac12,\pm\tfrac12)$ and $(1,\pm1)$, a feature clearly visible in Fig \ref{fig:3}.

  \item \textbf{Sum Rules:} The numerical values of our calculated magnetic moments satisfy both the Hao–Song sum rules \cite{Hao-Song sum rules} for the decuplet and the Coleman–Glashow sum rule \cite{sum1}\cite{sum2} for the octet representations. Specifically, the following relations are fulfilled for all $J^{P}$:

\begin{align*}
\mu_{P_{\psi \Delta}^+} - \mu_{P_{\psi}^0}
+ \mu_{P_{\psi s}^-} - \mu_{P_{\psi s}^+}
+ \mu_{P_{\psi ss}^0} - \mu_{P_{\psi ss}^-} &= 0,
\end{align*}
\begin{align*}
\mu_{P_{\psi \Delta}^{++}} + \mu_{P_{\psi \Delta}^-}
+ \mu_{P_{\psi s}^{-}} &= 3\,\mu_{P_{\psi s}^0},
\end{align*}

which correspond to the Hao–Song sum rules for the decuplet, and

\begin{equation*}
\mu_p - \mu_n + \mu_{\Sigma^-} - \mu_{\Sigma^+} + \mu_{\Xi^0} - \mu_{\Xi^-} = 0,
\end{equation*}

which represents the Coleman–Glashow sum rule for the baryon octet. The fulfillment of the magnetic-moment sum rules within each $SU(3)_F$ multiplet further supports the internal consistency of the constructed wave functions. 

  \end{itemize}
  Overall, the trends observed indicate that the magnetic moments of hidden-charm pentaquarks are governed primarily by their spin-flavor structure, with charge, strangeness, and isospin symmetry playing decisive roles in determining both the magnitude and the sign of the magnetic moments.

\begin{table*}[t]
\caption{Magnetic moments $\mu_B$ (in nuclear magnetons) of hidden-charm decuplet
pentaquark states for different spin--parity assignments and internal
spin couplings.}
\label{tab:decuplet_magnetic_moments}
\begin{ruledtabular}
\begin{tabular}{lccccccc}
State
& \multicolumn{3}{c}{$J^P=\tfrac12^-$}
& \multicolumn{3}{c}{$J^P=\tfrac32^-$}
& $J^P=\tfrac52^-$ \\
\cline{2-4}\cline{5-7}
& $\tfrac12^+\!\otimes0^-\!\otimes0^+$
& $\tfrac12^+\!\otimes1^-\!\otimes0^+$
& $\tfrac32^+\!\otimes1^-\!\otimes0^+$
& $\tfrac12^+\!\otimes1^+\!\otimes0^-$
& $\tfrac32^+\!\otimes0^-\!\otimes0^+$
& $\tfrac32^+\!\otimes1^-\!\otimes0^+$
& $\tfrac32^+\!\otimes1^-\!\otimes0^+$ \\
\hline
$P^{++}_{c}$    & 2.346  & 0.138  & 1.876  & 3.664  & 4.193  & 3.619  & 5.553 \\
$P^{+}_{c}$     & 1.084  & -0.065 & 1.164  & 1.529  & 2.363  & 1.911  & 2.808 \\
$P^{0}_{c}$     & -0.136 & -0.268 & 0.453  & -0.606 & 0.533  & 0.203  & 0.063 \\
$P^{-}_{c}$     & -1.356 & -0.471 & -0.259 & -2.741 & -1.297 & -1.505 & -2.682 \\
$P^{+}_{cs}$    & 1.203  & -0.045 & 1.234  & 1.738  & 2.542  & 2.078  & 3.076 \\
$P^{0}_{cs}$    & -0.016 & -0.248 & 0.522  & -0.397 & 0.712  & 0.370  & 0.331 \\
$P^{-}_{cs}$    & -1.236 & -0.451 & -0.189 & -2.532 & -1.118 & -1.338 & -2.414 \\
$P^{0}_{css}$   & 0.102  & -0.228 & 0.592  & -0.189 & 0.890  & 0.536  & 0.599 \\
$P^{-}_{css}$   & -1.117 & -0.432 & -0.120 & -2.324 & -0.940 & -1.172 & -2.146 \\
$P^{--}_{csss}$ & -0.998 & -0.412 & 0.050  & -2.115 & -0.761 & -1.005 & -1.878 \\
\end{tabular}
\end{ruledtabular}
\end{table*}

\begin{table*}[t]
\caption{Magnetic moments $\mu_B$ (in nuclear magnetons) of hidden-charm
octet $8_{1f}$ pentaquark states for different spin--parity assignments
and internal spin couplings.}
\label{tab:octet1f_magnetic_moments}
\begin{ruledtabular}
\begin{tabular}{lccccccc}
State
& \multicolumn{3}{c}{$J^P=\tfrac12^-$}
& \multicolumn{3}{c}{$J^P=\tfrac32^-$}
& $J^P=\tfrac52^-$ \\
\cline{2-4}\cline{5-7}
& $\tfrac12^+\!\otimes0^-\!\otimes0^+$
& $\tfrac12^+\!\otimes1^-\!\otimes0^+$
& $\tfrac32^+\!\otimes1^-\!\otimes0^+$
& $\tfrac12^+\!\otimes1^+\!\otimes0^-$
& $\tfrac32^+\!\otimes0^-\!\otimes0^+$
& $\tfrac32^+\!\otimes1^-\!\otimes0^+$
& $\tfrac32^+\!\otimes1^-\!\otimes0^+$ \\
\hline
$P^{+}_{c}$        & 1.694  & -0.878 & 1.978  & 1.224  & 3.278  & 2.216  & 2.808 \\
$P^{0}_{c}$        & -0.746 & 0.545  & -0.360 & -0.300 & -0.382 & -0.102 & 0.063 \\
$P^{+}_{cs}$       & 1.754  & -0.779 & 1.968  & 1.462  & 3.367  & 2.352  & 3.076 \\
$P^{0}_{cs}$       & 0.229  & -0.575 & 0.849  & -0.520 & 1.080  & 0.492  & 0.331 \\
$P^{-}_{cs}$       & -1.297 & -0.372 & -0.269 & -2.502 & -1.208 & -1.368 & -2.414 \\
$P^{0}_{css}$      & -0.448 & 0.505  & -0.142 & 0.086  & 0.064  & 0.261  & 0.599 \\
$P^{-}_{css}$      & -1.058 & -0.511 & -0.040 & -2.353 & -0.850 & -1.142 & -2.146 \\
$P^{0\prime}_{cs}$ & -0.262 & 0.079  & 0.195  & -0.274 & 0.343  & 0.247  & 0.331 \\
\end{tabular}
\end{ruledtabular}
\end{table*}

\begin{table*}[t]
\caption{Magnetic moments $\mu_B$ (in nuclear magnetons) of hidden-charm
octet $8_{2f}$ pentaquark states for different spin--parity assignments
and internal spin couplings.}
\label{tab:octet2f_magnetic_moments}
\begin{ruledtabular}
\begin{tabular}{lccc}
State
& \multicolumn{2}{c}{$J^P=\tfrac12^-$}
& $J^P=\tfrac32^-$ \\
\cline{2-3}
& $\tfrac12^+\!\otimes0^-\!\otimes0^+$
& $\tfrac12^+\!\otimes1^-\!\otimes0^+$
& $\tfrac12^+\!\otimes1^-\!\otimes0^+$ \\
\hline
$P^{+}_{c}$        & 0.491 & 0.743  & 1.851 \\
$P^{0}_{c}$        & 0.491 & -1.087 & -0.894 \\
$P^{+}_{cs}$       & 0.491 & 0.743  & 1.851 \\
$P^{0}_{cs}$       & 0.491 & -0.172 & 0.478 \\
$P^{-}_{cs}$       & 0.491 & -1.087 & -0.894 \\
$P^{0}_{css}$      & 0.491 & -0.908 & -0.626 \\
$P^{-}_{css}$      & 0.491 & -0.908 & -0.626 \\
$P^{0\prime}_{cs}$ & 0.491 & -0.663 & -0.258 \\
\end{tabular}
\end{ruledtabular}
\end{table*}

\section{Summary and Conclusions}

In this work, we have studied hidden-charm pentaquark states by combining mass
spectroscopy and magnetic-moment calculations within a molecular
baryon--meson picture. Using the Gürsey--Radicati (GR) mass model, we constrained
the possible spin--parity assignments of the experimentally observed $P_c$ and
$P_{cs}$ states and predicted the masses of their missing partners in the octet
and decuplet multiplets. Although the mass spectrum provides important
information on the multiplet structure, it is not sufficient by itself to fix
the spin--parity quantum numbers of multiquark states.

To obtain additional insight, we constructed quark-model wave functions and
calculated magnetic moments for different spin--parity assignments. The magnetic
moments show clear and systematic trends with total spin, charge, strangeness,
and internal spin--flavor symmetry. In particular, positively charged states
generally have larger magnetic moments for higher spin, with the largest values
appearing for $J^P=\tfrac52^-$. Negatively charged states show increasingly
negative magnetic moments for the same spin assignment, while neutral states
exhibit only a weak dependence on $J$ due to isospin cancellations. A gradual
reduction in the magnitude of the magnetic moments is also observed with
increasing strangeness.

When these magnetic-moment patterns are considered together with the GR mass
results, the ambiguity in assigning spin--parity quantum numbers to the observed pentaquark states is significantly reduced. The combined analysis favors the
assignments $P_c(4312)$ as $J^P=\tfrac12^-$, $P_c(4380)$ as $J^P=\tfrac32^-$,
$P_c(4440)$ and $P_c(4457)$ as $J^P=\tfrac52^-$, $P_{cs}(4338)$ as
$J^P=\tfrac12^-$ $(Y=0)$, and $P_{cs}(4458)$ as $J^P=\tfrac32^-$ $(Y=0)$. While several of these states are close in mass, their magnetic moments differ
qualitatively for alternative spin assignments, allowing some scenarios to be
favored over others.

It should be emphasized that experimental measurements of the magnetic moments
of these pentaquark states are not yet available. Therefore, the present
magnetic-moment results are not meant to provide direct experimental
confirmation. Instead, they offer complementary theoretical constraints that
help distinguish among different spin--parity possibilities that remain open
from mass spectroscopy alone.

In summary, the combined study of masses and magnetic moments provides a more
complete and useful picture of hidden-charm pentaquarks than spectroscopy by
itself. Our results support the interpretation of the observed $P_c$ and
$P_{cs}$ states are mainly octet pentaquarks and provide clear predictions for
their spin--parity assignments and electromagnetic properties. These predictions
can be tested in future experiments or lattice-QCD studies and may help clarify
the internal structure of exotic pentaquark states.

\end{document}